\documentclass[final,english]{bullsrsl}[2022/06/15]

\RequirePackage{geometry}

\usepackage[latin1]{inputenc}
\usepackage[T1]{fontenc}

\usepackage{siunitx}
\usepackage{natbib} 
\usepackage{graphicx}
\usepackage{gensymb}

\begin{document}
\title{A year-long representation of the ILMT observations in different coordinate systems}

\author[affil={1,2}, corresponding]{Monalisa}{Dubey}
\author[affil={1,3}]{Bhavya}{Ailawadhi}
\author[affil={4,5}]{Talat}{Akhunov}
\author[affil={6}]{Ermanno}{Borra}
\author[affil={1}]{Kuntal}{Misra}
\author[affil={1,2}]{Naveen}{Dukiya}
\author[affil={7}]{Jiuyang}{Fu}
\author[affil={7}]{Baldeep}{Grewal}
\author[affil={7}]{Paul}{Hickson}
\author[affil={1}]{Brajesh}{Kumar}
\author[affil={1,3}]{Vibhore}{Negi}
\author[affil={1,8}]{Kumar}{Pranshu}
\author[affil={7}]{Ethen}{Sun}
\author[affil={9}]{Jean}{Surdej}
\affiliation[1]{Aryabhatta Research Institute of observational sciencES (ARIES), Manora Peak, Nainital, 263001, India}
\affiliation[2]{Department of Applied Physics, Mahatma Jyotiba Phule Rohilkhand University, Bareilly, 243006, India}
\affiliation[3]{Department of Physics, Deen Dayal Upadhyaya Gorakhpur University, Gorakhpur, 273009, India}
\affiliation[4]{National University of Uzbekistan, Department of Astronomy and Astrophysics, 100174 Tashkent, Uzbekistan}
\affiliation[5]{Ulugh Beg Astronomical Institute of the Uzbek Academy of Sciences, Astronomicheskaya 33, 100052 Tashkent, Uzbekistan}
\affiliation[6]{Department of Physics, Universit\'{e} Laval, 2325, rue de l'Universit\'{e}, Qu\'{e}bec, G1V 0A6, Canada}
\affiliation[7]{Department of Physics and Astronomy, University of British Columbia, 6224 Agricultural Road, Vancouver, BC V6T 1Z1, Canada}
\affiliation[8]{Department of Applied Optics and Photonics, University of Calcutta, Kolkata, 700106, India}
\affiliation[9]{Institute of Astrophysics and Geophysics, University of Li\`{e}ge, All\'{e}e du 6 Ao$\hat{\rm u}$t 19c, 4000 Li\`{e}ge, Belgium}

\correspondance{monalisadubeyprl@gmail.com}
\date{11th September 2023}
\maketitle

\begin{abstract}
The 4m International Liquid Mirror Telescope (ILMT) is the first optical survey telescope in India that performs zenithal observations of a 22$'$ wide strip of the sky. To determine the portion of the sky covered by the ILMT during the entire year, we represent the ILMT Field of View (FoV) in three different coordinate systems - galactic, ecliptic, and equatorial. We adopt a constant declination of $+$\ang{29;21;41.4} and varying right ascension (RA) ranges corresponding to the Local Sidereal Time (LST). The observations from June to September are hampered due to the monsoon season. The handiness of such representations will allow us to locate a transient event in the ILMT FoV. This will enable prompt follow-up observations with other facilities.
\end{abstract}


\section{Introduction}
\label{introduction}

The 4m International Liquid Mirror Telescope (ILMT), due to its liquid mercury mirror, is designed to continuously scan a 22$'$ wide strip of the zenith sky in g', r', and i' filters in Time Delay Integration (TDI) mode with a 4K $\times$ 4K CCD camera. A detailed description of the telescope is presented in \citet{Surdej2018}. The telescope achieved first light on April 29 2022 \citep{BKumar2022} and has been undergoing the commissioning phase since then.

The ILMT is Asia's largest liquid mirror telescope located at the Devasthal Observatory of ARIES, India (longitude \ang{79;41;07.08} East and latitude $+$\ang{29;21;41.4} North) at an elevation of 2378$\pm 20$ m. It is the first dedicated optical survey telescope in India. Due to its strategic location between Eastern Australia and the Canary Islands, it is best suited for discovering new astronomical sources (mainly transients such as supernovae). In this work, we present the ILMT Field of View (FoV) in three different coordinate systems: 1) galactic, 2) ecliptic, and 3) equatorial for a ready reference. The representation of the FoV is given in Section\,\ref{coordinates} followed by a brief summary in Section\,\ref{summary}.

\section{The ILMT Field of View over a year}
\label{coordinates}

Astronomical coordinate systems are essential for determining the location of any astronomical object on the celestial sphere. There are mainly four coordinate systems: horizontal, equatorial, ecliptic, and galactic. We map the ILMT FoV in equatorial, ecliptic, and galactic coordinate systems to comprehend the position of celestial objects that we can track with this telescope. The angular area of the sky extent through a telescope is referred to as its FoV.  Depending on the coordinate system used to characterise the observations, a wealth of information on the celestial sources observed can be inferred. 

In the equatorial coordinate system, the right ascension (RA) and declination of celestial objects within the FoV are secured. This helps to identify and catalog the stars, galaxies, and other astronomical objects such as transients. This information further contributes to astrometry, the science of celestial positioning. By observing the same region of the sky each night, we can measure the apparent motion of stars and determine their proper motion. 

The FoV in the ecliptic coordinate system is useful for studying the objects within the solar system, especially those that lie close to the ecliptic plane. The FoV of the ecliptic coordinate system allows to study the motion of planets, asteroids, comets, and other objects in the Solar System along their orbits. By following them, we can determine not only their apparent motion, orbital elements, and future positions but also their composition and origins. By observing and analysing the intensity, distribution, and polarisation of sunlight within this FoV, we can gain insights into the properties of interplanetary dust, the structure of the zodiacal cloud, and the formation processes in the solar system. Hence, the ecliptic coordinate system provides us with an opportunity to study the dynamics, interactions, and phenomena associated with the solar system. 

The trace of the telescope's sky route with regard to the galactic plane is provided by the FoV of the galactic coordinate system. By observing stars, clusters, and nebulae within the FOV, we can determine their position, which gives an idea of the distribution and characteristics of objects within the Milky Way galaxy. In this way, we can gain insights into the overall motion and behaviour of the galaxy.

\begin{figure}[t]
\centering
\includegraphics[width=\textwidth]{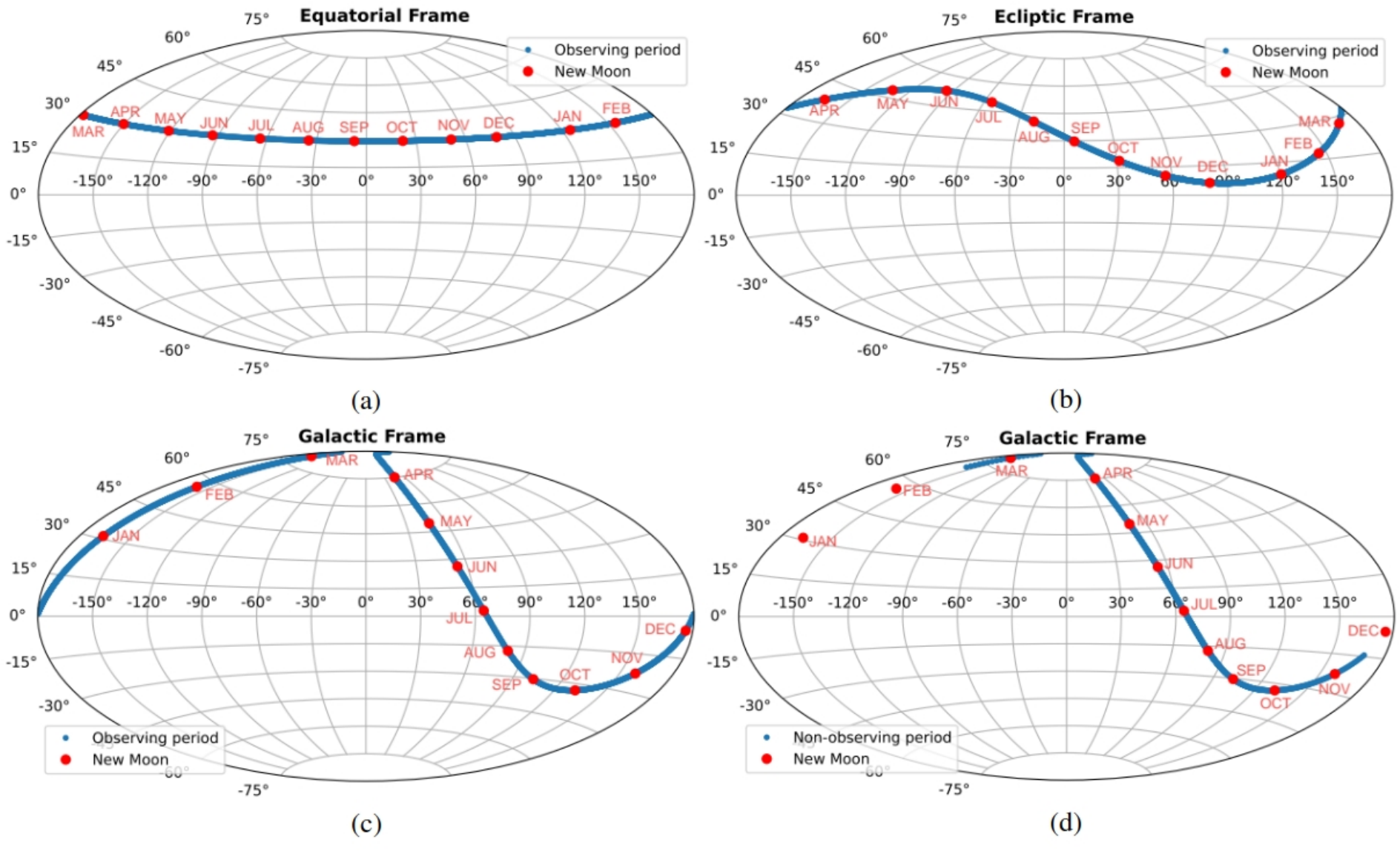}
\bigskip

\begin{minipage}{120mm}
\caption{Representation of the ILMT FoV for the year 2023 during the observational period between October--May in the (a) equatorial, (b) ecliptic, and (c) galactic coordinate systems. The bottom right figure (d) represents the ILMT FoV in the galactic coordinate system during the non-observing (June--September) monsoon period. The RA ranges in the non-observing period overlap with the RA ranges covered during the observing period indicating that all RA ranges are observable with the ILMT from October--May. The red dots in the figure indicate the new moon day in each month for the year 2023.}
\label{ILMT_FoV}
\end{minipage}
\end{figure}

As ILMT performs zenith sky observations at constant declination centered at $+$\ang{29;21;41.4} (the latitude of Devasthal Observatory), the source moves in the focal plane along the RA axis that corresponds to the LST at the time of transit.

The primary goal of this telescope is to detect transients (such as supernovae), asteroids, and variable stars. A year-long representation of ILMT's FoV in galactic coordinates is advantageous in locating the celestial source because we already have the information on the patch of the sky being surveyed by the ILMT at that particular time. 

\begin{table}
\centering
\begin{minipage}{160mm}
\caption{List of LST (at the Devasthal Observatory site) on each new moon date of the observing period (October--May) along with the corresponding galactic and ecliptic longitude (Lon) and latitude (Lat).}
\end{minipage}
\bigskip

\setlength\tabcolsep{4.0pt}
\begin{tabular}{crrrrr}
\hline
\textbf{Date}         & \multicolumn{1}{c}{\textbf{LST}}    & \multicolumn{1}{c}{\textbf{Galactic Lon}} & \multicolumn{1}{c}{\textbf{Galactic Lat}} & \multicolumn{1}{c}{\textbf{Ecliptic Lon}} & \multicolumn{1}{c}{\textbf{Ecliptic Lat}} \\
\textbf{(yyyy-mm-dd)} & \multicolumn{1}{c}{\textbf{(hour)}} & \multicolumn{1}{c}{\textbf{(\si{\degree} \hspace{0.1cm} \si{\arcminute} \hspace{0.1cm} \si{\arcsecond})}} & \multicolumn{1}{c}{\textbf{(\si{\degree} \hspace{0.1cm} \si{\arcminute} \hspace{0.1cm} \si{\arcsecond})}} & \multicolumn{1}{c}{\textbf{(\si{\degree} \hspace{0.1cm} \si{\arcminute} \hspace{0.1cm} \si{\arcsecond})}} & \multicolumn{1}{c}{\textbf{(\si{\degree} \hspace{0.1cm} \si{\arcminute} \hspace{0.1cm} \si{\arcsecond})}} \\
\textbf{Time}                  \\
\textbf{(hh:mm:ss)}            \\
\hline
2023-01-21 &  2.86 &    \ang{151;51;11.16} & $-$\ang{26;51;47.13} &    \ang{ 49;08;15.97} & \ang{12;22;02.48} \\
19:00:00 \\
2023-01-22 & 11.88 & $-$\ang{160;44;20.31} &    \ang{76;38;34.75} &    \ang{165;49;33.96} & \ang{26;02;24.16} \\
04:00:00 \\
2023-02-20 &  4.83 &    \ang{172;26;08.74} & $-$\ang{10;01;42.10} &    \ang{ 74;37;03.65} & \ang{06;51;34.43} \\
19:00:00 \\
2023-02-21 & 13.85 &    \ang{ 46;39;53.91} &    \ang{76;53;22.75} & $-$\ang{167;01;41.22} & \ang{37;41;19.80} \\
04:00:00 \\
2023-03-21 &  6.73 & $-$\ang{174;30;39.26} &    \ang{11;09;43.04} &    \ang{ 99;37;39.48} & \ang{06;17;28.92} \\
19:00:00 \\
2023-03-22 & 15.76 &    \ang{ 46;52;13.28} &    \ang{52;02;28.22} & $-$\ang{135;43;52.50} & \ang{47;35;49.44} \\
04:00:00 \\
2023-04-19 &  8.64 & $-$\ang{165;34;10.70} &    \ang{34;38;14.94} &    \ang{124;23;18.96} & \ang{10;31;32.45} \\
19:00:00 \\
2023-04-20 & 17.66 &    \ang{ 53;37;35.77} &    \ang{27;42;47.76} & $-$\ang{ 97;17;09.21} & \ang{52;40;06.61} \\
04:00:00 \\
2023-05-19 & 10.61 & $-$\ang{159;40;50.97} &    \ang{60;03;02.40} &    \ang{149;30;16.10} & \ang{19;03;10.71} \\
19:00:00 \\
2023-05-20 & 19.63 &    \ang{ 63;58;33.24} &    \ang{03;58;52.07} & $-$\ang{ 55;49;54.47} & \ang{49;55;52.06} \\
04:00:00 \\
2023-10-14 & 20.33 &    \ang{ 68;45;02.28} & $-$\ang{03;49;53.80} & $-$\ang{ 42;36;13.49} & \ang{47;12;10.20} \\
19:00:00 \\
2023-10-15 &  5.36 &    \ang{176;37;35.40} & $-$\ang{04;27;00.34} &    \ang{ 81;35;18.42} & \ang{06;12;18.50} \\
04:00:00 \\
2023-11-13 & 22.31 &    \ang{ 86;51;31.75} & $-$\ang{22;37;31.31} & $-$\ang{ 10;39;32.66} & \ang{36;46;19.93} \\
19:00:00 \\
2023-11-14 &  7.33 & $-$\ang{171;21;01.80} &    \ang{18;21;19.12} &    \ang{107;27;04.78} & \ang{07;07;32.90} \\
04:00:00 \\
2023-12-12 &  0.21 &    \ang{112;33;51.41} & $-$\ang{32;49;57.54} &    \ang{ 15;22;26.59} & \ang{25;30;41.62} \\
19:00:00 \\
2023-12-13 &  9.24 & $-$\ang{163;23;18.65} &    \ang{42;15;52.66} &    \ang{132;02;37.72} & \ang{12;43;50.95} \\
04:00:00 \\
\hline
\end{tabular}
\label{Tab:Table 1}
\end{table}

A sample list of LST ranges (at the Devasthal Observatory site) for the year 2023 on new moon dates with the corresponding galactic and ecliptic coordinates for the start and end times of night observations is given in Table\,1. This instantly provides the RA ranges being covered on a particular day at a given time. Due to the monsoon season, observations from June to September (4 months) are hampered at the Devasthal Observatory. Fig.\,\ref{ILMT_FoV} represents the ILMT FoV in equatorial, ecliptic, and galactic coordinate systems for the observing season between October--May. In the equatorial plane, the FoV is like a strip of varying RA ranges with a constant declination ($+$\ang{29;21;41.4}). In the ecliptic frame, the sky path of the ILMT is constrained only by positive latitude. The monsoon season's sky route is depicted in the bottom right plot. Since during a single night, the ILMT typically covers the LST range between 8 and 12 hours, corresponding to 120\degree--180\degree, the most extreme RA reached on June 1 at the end of the night does in fact correspond to the less extreme RA reached at the beginning of the night on October 1. Indeed, the change in LST during a period of 4 months at the start of a night does correspond to approximately 120\degree. 

\section{Summary}
\label{summary}

The ILMT is designed to perform deep imaging in $i\text{'}$, $r\text{'}$, and $g\text{'}$ bands of the 22$'$ wide strip of the sky each night in TDI mode. A particularly advantageous opportunity to find and monitor transients like supernovae, asteroids, and other variable stars will arise from the one-day cadence. With a representation of the ILMT FoV, one can list the RA range and the annual sky coverage with the ILMT. Knowing the ILMT FoV in advance not only helps to anticipate the rate at which astronomical events will be detected but also spells out the area of the sky we will view with the telescope. 

\begin{acknowledgments}
The 4m International Liquid Mirror Telescope (ILMT) project results from a collaboration between the Institute of Astrophysics and Geophysics (University of Li\`{e}ge, Belgium), the Universities of British Columbia, Laval, Montreal, Toronto, Victoria and York University, and Aryabhatta Research Institute of observational sciencES (ARIES, India). The authors thank Hitesh Kumar, Himanshu Rawat, Khushal Singh and other observing staff for their assistance at the 4m ILMT.  The team acknowledges the contributions of ARIES's past and present scientific, engineering and administrative members in the realisation of the ILMT project. JS wishes to thank Service Public Wallonie, F.R.S.-FNRS (Belgium) and the University of Li\`{e}ge, Belgium for funding the construction of the ILMT. PH acknowledges financial support from the Natural Sciences and Engineering Research Council of Canada, RGPIN-2019-04369. PH and JS thank ARIES for hospitality during their visits to Devasthal. B.A. acknowledges the Council of Scientific $\&$ Industrial Research (CSIR) fellowship award (09/948(0005)/2020-EMR-I) for this work. M.D. acknowledges Innovation in Science Pursuit for Inspired Research (INSPIRE) fellowship award (DST/INSPIRE Fellowship/2020/IF200251) for this work. T.A. thanks Ministry of Higher Education, Science and Innovations of Uzbekistan (grant FZ-20200929344). 
\end{acknowledgments}

\newpage

\begin{furtherinformation}

\begin{orcids}
\orcid{0000-0003-1637-267X}{Kuntal}{Misra}
\orcid{0000-0002-7005-1976}{Jean}{Surdej}
\end{orcids}

\begin{authorcontributions}
This work results from a long-term collaboration to which all authors have made significant contributions.

\end{authorcontributions}

\begin{conflictsofinterest}
The authors declare no conflict of interest.

\end{conflictsofinterest}

\end{furtherinformation}

\bibliographystyle{bullsrsl-en}

\bibliography{S11-P11_DubeyM}

\end{document}